\documentclass[10pt]{article}
\usepackage{amsmath}
\usepackage{graphicx}
\usepackage{color}
\usepackage{orcidlink}
\usepackage{hyperref}
\hypersetup{colorlinks=true, linkcolor=blue, citecolor=blue, urlcolor=blue}
\usepackage{caption}
\usepackage{subcaption}
\begin{document}
\baselineskip=12 pt

\begin{center}
{\large {\bf Quantum motions of scalar particles under rainbow gravity effects in spherical symmetrical G\"{o}del-type metric with topological defects }}
\end{center}

\vspace{0.2cm}

\begin{center}
    {\bf Faizuddin Ahmed\orcidlink{0000-0003-2196-9622}}\footnote{\bf faizuddinahmed15@gmail.com ; faizuddin@ustm.ac.in}\\
    \vspace{0.1cm}
    {\it Department of Physics, University of Science \& Technology Meghalaya, Ri-Bhoi, 793101, India}\\
    \vspace{0.2cm}
    {\bf Abdelmalek Bouzenada\orcidlink{0000-0002-3363-980X}}\footnote{\bf abdelmalekbouzenada@gmail.com ; abdelmalek.bouzenada@univ-tebessa.dz }\\
    \vspace{0.1cm}
    {\it  Laboratory of Theoretical and Applied Physics, Echahid Cheikh Larbi Tebessi University, Algeria}\\
\end{center}

\vspace{0.2cm}
\begin{abstract}
In this research contribution, we explore the relativistic quantum dynamics of spin-0 scalar particles within a curved space-time background containing topological defects, while incorporating the effects of rainbow gravity. The Klein–Gordon equation is solved in a spherically symmetrical G\"{o}del-type space-time, considering the presence of topological defects and the influence of rainbow gravity, described by the line-element: $ds^{2}=-\left\{\frac{dt}{\mathcal{F}}+\alpha\,\Omega\,r^{2}\Big(1+\frac{r^{2}}{4\,R^{2}}\Big)^{-1}\frac{d\varphi}{\mathcal{G}}\right\}^{2} +\mathcal{G}^{-2}\,\left\{\left(1+\frac{r^{2}}{4R^{2}}\right)^{-2}\left(dr^{2}+\alpha^{2}r^{2} d\varphi^{2}\right)+dz^{2}\right\}$. The functions $\mathcal{F}=\mathcal{F}(\chi)$ and $\mathcal{G}=\mathcal{G}(\chi)$, where $0 < \chi(=\frac{|E|}{E_p}) \leq 1$, $E$ is the particle's energy, and $E_p$ the Planck's energy, are referred to as rainbow functions. Analytical solutions for the energy levels and wave-functions of scalar particles are obtained through general hypergeometric equations for the pair of rainbow functions: (i) $\mathcal{F}(\chi)=1$ and $\mathcal{G}(\chi)=(1+\kappa\,\chi)$, and (ii) $\mathcal{F}(\chi)=\frac{(e^{\beta_0\,\chi}-1)}{\beta_0\,\chi}$ and $\mathcal{G}(\chi)=1$. Notably, we demonstrate that the energy levels and wave-functions of scalar particles are intricately linked to the topological defect of cosmic string, characterized by the parameter $\alpha$, and the constant angular speed of the rotating frame, characterized by the parameter $\Omega$. Furthermore, the influence of the rainbow parameters $(\kappa, \beta_0)$ is evident in the eigenvalue solutions, introducing modifications compared to known expressions. It is worth mentioning that the presence of topological defects breaks the degeneracy of the energy levels, resulting in substantial modifications. These findings underscore the interplay between rainbow gravity, topological defects, and the intrinsic properties of scalar particles within a curved space-time framework.
\end{abstract}

\vspace{0.1cm}


\vspace{0.1cm}


\section{Introduction}

The unification of quantum theory and general relativity into a theory of 'quantum gravity' remains one of the great unresolved problems in theoretical physics. Although general relativity successfully describes cosmological phenomena, and quantum field theory is highly accurate in particle physics, attempts to quantize gravity encounter infinities, making general relativity a non-renormalizable theory. To address this challenge, various approaches, such as Loop Quantum Gravity and Supersymmetry, have been proposed. Quantum field theory in curved space-time is an initial attempt to grapple with the quantization of gravity. However, analyzing phenomena where quantum and gravitational effects are crucial can be done within the semiclassical theory of quantum field theory on curved space-time. This theory operates in the intermediate range between the Planck scale (where quantum effects become significant, approximately $10^{-35}$m) and the standard model scale (approximately $10^{-19}$m). In this range, a theory of quantum fields propagating on a classical curved space-time background can be employed. For larger length scales, the semiclassical approximation of quantum gravity is often sufficient. This involves investigating quantum fields on curved space-times, ensuring that the semi-classical Einstein field equations $G_{\mu\nu}=T_{\mu\nu}$ are satisfied. However, the transition from flat Minkowski space to a curved space poses challenges for quantum field theory (QFT). Curved space-times lack Poincaré symmetry, eliminating the possibility of a global Fourier transformation. Consequently, field expansions into positive and negative frequency solutions become problematic. The absence of Poincaré invariance also leads to difficulties in defining a unique vacuum state within a curved space-time. In the Wightman formalism, the vacuum state is explicitly defined by its Poincaré invariance and the spectrum condition. Due to the lack of a unique vacuum state, establishing a unique Fock space becomes challenging, and describing particle creation and annihilation becomes problematic. To address these curved space-time issues, the algebraic approach to QFT proves to be a more suitable framework. For a comprehensive review of quantum field theory in curved space, readers are referred to \cite{NDB, SAF, RMW, LEP}. 

The early universe may have given rise to various topological defects \cite{V1, V2}, such as domain walls \cite{k1}, global monopoles \cite{k2}, and cosmic strings \cite{M20, V3, V4, V5, V6, V7}, during the vacuum phase transition. Cosmic strings, in particular, have garnered significant attention due to their impact on the medium's topology and their association with the emergence of topological defects in space-time. From a field-theoretical perspective, cosmic strings are hypothetical one-dimensional topological defects that could emerge during the symmetry-breaking phase transition in the early universe. The gravitational field around a cosmic string is intriguing, as a particle at rest near an infinite, static cosmic string does not experience attraction, leading to a lack of local gravity. Although the space-time surrounding the string is locally flat, it globally exhibits a conical topology. For a comprehensive review of cosmic string and other topological defects, readers are as refereed as \cite{RB1, RB2, RB3, RB4, RB5, RB6} and for experimental status see Ref. \cite{Planck 2013, epjc}.  

In 1949, Gödel \cite{key-1} derived a solution from Einstein's equations, known as the first cosmological solution for rotating matter. Noteworthy characteristics of this solution include its stationary nature, spatial homogeneity, and cylindrical symmetry. One of its intriguing properties is the violation of causality, introducing the possibility of closed time-like curves (CTCs). It is essential to note that the presence of CTCs, as speculated by Hawking \cite{key-2}, is considered physically inconsistent. The metric underwent generalization in cylindrical coordinates in \cite{key-3, key-4, key-5}, where the causality issue was scrutinized in greater detail. Consequently, three distinct classes of solutions emerged, each characterized by specific possibilities: (i) the absence of CTCs, (ii) an infinite sequence of alternating causal and non-causal regions, and (iii) the presence of only one non-causal region. The paper \cite{key-6} introduced the concept of super-energy and super-momentum as criteria for determining the potential existence of CTCs in \cite{key-7, key-8}. Another fact contributing to the interest in the Gödel solution is its allowance for non-trivially embedded black holes \cite{key-10}. Diverse aspects of the Gödel solutions are explored in \cite{key-11, key-12, key-13, key-14}. A recent study \cite{M1} demonstrated that Gödel-type metrics with flat backgrounds are precise solutions to the Einstein-Aether theory in four dimensions \cite{M2}. These space-times have been explored in both Riemannian Gödel-type space-times \cite{key-3} and Riemann-Cartan Gödel-type space-times \cite{M12, M13, M14}. Various aspects of Gödel solutions are also examined in \cite{M15, M16, M17, M18}, and their implications in string theory are discussed in \cite{M19}.  

The family of Gödel-type solutions has sparked extensive investigations into geodesics within these solutions across various physical contexts, as evident in references \cite{MM22, MM23, MM24, MM25, MM26}. Recently, the spherical Gödel-type solution has found application in exploring electronic properties within a geometric model describing the fullerene molecule \cite{MM27}. The quantum dynamics of scalar particles and spin-1/2 particles in Gödel-type space-times have been subjects of exploration by several researchers. The initial examination of this issue was conducted in \cite{MM22}, where the Klein-Gordon and Dirac equations in Gödel-type space-time with both positive and negative curvatures, and in flat Gödel-type space-time was investigated. Drukker {\it et al.} \cite{MM23} investigated the connection between quantum dynamics of a scalar particle in the four-dimensional class of Gödel solutions in general relativity and the Landau levels in flat, spherical, and hyperbolic spaces. Their study explored the similarity between the energy levels of the scalar quantum particle in Gödel-type space-times, considering backgrounds with zero, positive, and negative curvatures. This analogy was further observed in \cite{MM28}, where quantum dynamics of scalar particles in the Som-Raychaudhuri space-time (Gödel flat solution) was investigated and compared the results with Landau levels in flat space. A few other investigations of the relativistic wave equations, such as the Dirac equation and its oscillator, the Duffin-Kemmer-Petiau (DKP) equation, the Klein-Gordon equation and its oscillator have been investigated in G\"{o}del and G\'{o}del-type metric backgrounds without and with topological defects by numerous authors in quantum system (see, examples in \cite{hhh1, hhh2, hhh3, hhh4, hhh5, hhh6, MM29, MM30} and related references there in). The G\"{o}del-type metrics in cylindrical coordinates $(t, r, \varphi, z)$ in the presence of topological defect of cosmic string can be expressed by the following line-element\cite{key-3, key-4, key-5, hhh1, hhh2, hhh3, hhh4, hhh5, hhh6, MM29, MM30}: 
\begin{equation}
    ds^{2}=-\left(dt+\alpha\,\Omega\, r^{2}\,\frac{\sinh^2 (\ell\,r)}{\ell^2}\,d\varphi\right)^{2}+\alpha^{2}\,\frac{\sinh^2 (2\,\ell\,r)}{4\,\ell^2}\,d\varphi^{2}+dr^2+dz^{2},\label{a1}
\end{equation}
where the variables $(r, \varphi, z, t)$ can take, respectively, the following values: $0 \leq r < \infty$, $0 \leq \varphi < 2\,\pi$, $-\infty < (z, t) < \infty$. The parameter $\Omega$ characterizes the vorticity of the space and $\alpha$ is the topological defect parameter. In recent years, the exploration of the quantum dynamics of particles in the presence of a uniform magnetic field has expanded to various types of curved spaces. Noteworthy studies include investigations of the Landau problem in hyperbolic space \cite{MM31,MM32}, the one-electron atom problem in topological defect \cite{M22}, and various curved space-time backgrounds in the presence of topological defects in Refs. \cite{MM33, k4, k5, ss1, ss2, ss3, ss6, ss5, ss6, ss11, ss12}.

Rainbow gravity, a framework exploring high-energy phenomena in quantum gravity, introduces higher-order terms in the energy-momentum dispersion relation through rainbow functions. This leads to a breakdown of Lorentz symmetry at the Planck energy scale \cite{KM1, KM2, KM3}. Rainbow functions govern deviations from conventional relativistic expressions, necessitating an energy-dependent space-time metric, especially as the probing particle's energy approaches the Planck scale. As an effective theory of quantum gravity, rainbow gravity incorporates Doubly Special Relativity principles, maintaining invariant energy and length scales alongside the speed of light, notably at lower energies \cite{KM1, KM2, KM4, KM5}. Its significance lies in modifying classical theories, such as integrating with Friedmann–Robertson–Walker cosmologies to explore resolutions for the Big Bang singularity \cite{KM6} and investigating inflationary solutions in modified models \cite{KM7}. Rainbow gravity extends its reach to phenomena like light deflection, gravitational red-shift, and weak equivalence principles \cite{KM8}. Recent investigations focused into relativistic particle behaviors in diverse space-times within rainbow gravity, exploring phenomena such as Landau levels, Dirac oscillators, and the quantum dynamics of photons \cite{KM9, KM10, KM11, KM12, KM13}. Additionally, studies have examined the relativistic motions of scalar and oscillator fields in various curved space-time backgrounds without and with topological defect in the literature \cite{ss7, ss8, ss9, ss10, KM14, KM15}.

Our aim to investigate the influence of rainbow gravity's on relativistic quantum dynamics of spin-0 scalar particles in a curved space-time background. The topological defect of cosmic string is introduced in this curved space-time, specifically in spherical symmetrical Gödel-type metric. We derive the radial wave equation and after a few mathematical steps convert into the general hypergeometric equation form. For suitable pair of rainbow functions, we obtain the energy levels and the wave-functions of scalar particles and analyze the results. We show that the energy levels and wave functions get modification by the rainbow parameter. Moreover, the presence of topological defect modify the eigenvalue solution and breaks the degeneracy of energy spectrum. This paper is structured as follows: in {\it section 2}, we write down a spherical symmetrical G\"{o}del-type metric with cosmic string in the presence of rainbow gravity where the temporal part is replaced by $dt \to \frac{dt}{\mathcal{F}}$ and the spatial part by $dx^{i} \to \frac{dx^i}{\mathcal{G}}$, where $\mathcal{F}=\mathcal{F}(\chi), \mathcal{G}=\mathcal{G}(\chi)$ are rainbow functions. Then we derive the radial equation of the Klein-Gordon equation in this geometry background and solve the equation through general hypergeometric function. We choose two pair of rainbow functions and present the eigenvalue solutions of the scalar particle. In {\it section 3}, we present our results and discussion.   

\section{Scalar particle in spherical symmetrical Gödel space-time with topological defects under rainbow gravity's}

Now, we consider the limit of (\ref{a1}) where we can obtain a class of solutions of Gödel type possessing spherical symmetry. We suggest that $\ell^2 <0$ and introduce the new coordinates in (\ref{a1}): $R=\frac{r}{2\,\ell}$ and $\theta=r/R$. In the case when the sign of $\ell^2 <0$, the metric (\ref{a1}) takes the following form \cite{MM29}:
\begin{equation}
ds^{2}=-\left(dt+\frac{\alpha\,\Omega\,r^{2}}{1+\frac{r^{2}}{4\,R^{2}}}\,d\varphi\right)^{2}+\left(1+\frac{r^{2}}{4R^{2}}\right)^{-2}\, \left(dr^{2}+\alpha^{2}\,r^{2}\,d\varphi^{2}\right)+dz^{2}\,.\label{b1}
\end{equation}

Now, we introduce rainbow gravity's in this space-time by modifying the temporal part $dt \to \frac{dt}{\mathcal{F}}$ and spatial part $dx^i \to \frac{dx^i}{\mathcal{G}}$, where $\mathcal{F}=\mathcal{F}(\chi)$ and $\mathcal{G}=\mathcal{G}(\chi)$ are referred to as rainbow functions. Here $0 < \chi(=\frac{|E|}{E_p}) \leq 1$ with $E$ is the particle's energy, and $E_p$ the Planck's energy. Therefore, space-time (\ref{b1}) under rainbow gravity becomes
\begin{equation}
ds^{2}=-\left(\frac{dt}{\mathcal{F}}+\frac{\alpha\,\Omega\,r^2}{1+\frac{r^{2}}{4\,R^{2}}}\,\frac{d\varphi}{\mathcal{G}}\right)^{2}+\left(1+\frac{r^{2}}{4R^{2}}\right)^{-2}\,\mathcal{G}^{-2}\,\left(dr^2+\alpha^2\,r^2\,d\varphi^2\right)+\frac{dz^2}{\mathcal{G}^2}\,.\label{b2}
\end{equation}

One can see that in the limit $R \to \infty$, the above deformed curved space-time reduces to the following form
\begin{equation}
ds^{2}=-\left(\frac{dt}{\mathcal{F}}+\alpha\,\Omega\,r^2\,\frac{d\varphi}{\mathcal{G}}\right)^{2}+\mathcal{G}^{-2}\,\left(dr^2+\alpha^2\,r^2\,d\varphi^2+dz^2\right)\label{bb2}
\end{equation}
which is the deformed metric form for the Som-Raychaudhuri line-element with topological defect. 

The covariant and contravariant metric tensor for the line-element (\ref{b2}) are given by
\begin{equation}
    g_{\mu\nu}=\begin{pmatrix}
    -\frac{1}{\mathcal{F}^2} & 0 & -\frac{1}{\mathcal{F}\,\mathcal{G}}\,\left(\frac{\alpha\,\Omega\, r^{2}}{1+\frac{r^{2}}{4\,R^{2}}}\right) & 0\\
    0 & \mathcal{G}^{-2}\,\left(1+\frac{r^{2}}{4\,R^{2}}\right)^{-2} & 0 & 0\\
    -\frac{1}{\mathcal{F}\,\mathcal{G}}\,\left(\frac{\alpha\,\Omega\, r^{2}}{1+\frac{r^{2}}{4\,R^{2}}}\right) & 0 & \frac{\alpha^{2}\,r^{2}\,(1-\Omega^2\,r^2)}{\mathcal{G}^2\,\Big(1+\frac{r^{2}}{4\,R^{2}}\Big)^{2}} & 0\\
    0 & 0 & 0 & \frac{1}{\mathcal{G}^2}
\end{pmatrix},\nonumber\\
\end{equation}
\begin{equation}
    g^{\mu\nu}=\left(\begin{array}{cccc}
    \mathcal{F}^2\,(-1+\Omega^2\,r^2) & 0 & -\frac{\Omega}{\alpha}\,\mathcal{F}\,\mathcal{G}\,\left(1+\frac{r^2}{4\,R^2}\right) & 0\\
    0 & \mathcal{G}^2\,\left(1+\frac{r^2}{4\,R^2}\right)^2 & 0 & 0\\
    -\frac{\Omega}{\alpha}\,\mathcal{F}\,\mathcal{G}\,\left(1+\frac{r^2}{4\,R^2}\right) & 0 & \frac{\mathcal{G}^2}{\alpha^2\,r^2}\,\left(1+\frac{r^2}{4\,R^2}\right) & 0\\
    0 & 0 & 0 & \mathcal{G}^2
\end{array}\right)\,.\label{b3}
\end{equation}
The determinant of the metric tensor $g_{\mu\nu}$ is given by
\begin{equation}
\sqrt{-g}=\frac{\alpha\,r}{\mathcal{F}\,\mathcal{G}^3\,\left(1+\frac{r^2}{4\,R^2}\right)^2}\,.\label{b4}
\end{equation}

Now we will investigate a scalar quantum particle in a Gödel type space-time under the influence of rainbow gravity's. The relativistic quantum dynamics of a free spinless particle of mass M is described by the Klein–Gordon equation. In its covariant form, this equation takes the following form ($\hbar=c=G=1$) \cite{ss1, ss2, ss3, ss4, ss5, ss6, ss11, ss12, ss7, ss8, ss9, ss10}:
\begin{equation}
    \Big[-\frac{1}{\sqrt{-g}}\,\partial_{\mu}\,(\sqrt{-g}\,g^{\mu\nu}\,\partial_{\nu})+M^2\Big]\,\Psi (t, \vec{r})=0,\label{b5}
\end{equation}
with $g$ being the determinant of the metric tensor $g_{\mu\nu}$ with its inverse $g^{\mu\nu}$. The $\Psi (t, \vec{r})$ is the amplitude of the probability to find the particles around the $\vec{r}$ position at the time $t$. 

Expressing the wave equation (\ref{b5}) in the space-time background (\ref{b2}) and using (\ref{b3})-(\ref{b4}), we get the following equation
\begin{eqnarray}
&&\Bigg[\mathcal{G}^2\,\left(1+\frac{r^{2}}{4\,R^{2}}\right)^2\,\frac{1}{r}\,\frac{\partial}{\partial r}\,\Big(r\,\frac{\partial}{\partial r}\Big)+\frac{\mathcal{G}^2}{\alpha^2\,r^2}\,\frac{\partial^2}{\partial \varphi^2}+\frac{r^2}{16\,R^4\,\alpha^2}\,\Big(\mathcal{G}\,\frac{\partial}{\partial \varphi}-4\,\alpha\,\Omega\,R^2\,\mathcal{F}\,\frac{\partial}{\partial t}\Big)^2-\mathcal{F}^2\,\frac{\partial^2}{\partial t^2}\nonumber\\
&&-\mathcal{G}^2\,\frac{\partial^2}{\partial z^2}-\frac{2\,\Omega\,\mathcal{F}\,\mathcal{G}}{\alpha}\,\frac{\partial}{\partial t}\,\frac{\partial}{\partial \varphi}+\frac{\mathcal{G}^2}{2\,\alpha^2\,R^2}\,\frac{\partial^2}{\partial \varphi^2}-M^2  \Bigg]\,\Psi=0\,.\label{b7}
\end{eqnarray}

It is easy to see that this differential equation also has a translation symmetry along the z-axis and an azimuthal symmetry. This allows us once again to write the solution as
\begin{equation}
\Psi (t, r, \varphi, z)=e^{-i\,E\,t}\,e^{i\,m\,\varphi}\,e^{i\,k_{z}\,z}\,\psi (r),\label{b8}
\end{equation}
where $E$ is the particle's energy, $m$ is the magnetic quantum number, $k$ is an arbitrary constant, and $\psi (r)$ is the radial wave function.

Substituting this solution in the above equation (\ref{b7}) results the following differential equation:
\begin{eqnarray}
    \left(1+\frac{r^{2}}{4\,R^{2}}\right)^2\,\Big(\psi''(r)+\frac{1}{r}\,\psi'(r)\Big)+\Bigg[\Lambda-\omega^2\,\,r^2-\frac{j^2}{r^2}\Bigg]\,\psi(r)=0,\label{b9}
\end{eqnarray}
where
\begin{eqnarray}
\Lambda=\frac{\mathcal{F}^2\,E^2-M^2}{\mathcal{G}^2}-k^2_{z}-\frac{2\,\Omega\,m\,E}{\alpha}\,\frac{\mathcal{F}}{\mathcal{G}}-\frac{m^2}{2\,\alpha^2\,R^2},\quad j=\frac{|m|}{\alpha},\quad \omega=\frac{b}{4\,R^2},\quad  b=\Big|\frac{m}{\alpha}+4\,\Omega\,R^2\,E\,\frac{\mathcal{F}}{\mathcal{G}}\Big|.\label{b10}
\end{eqnarray}
We introduce the change of variables $r=2\,R\,\tan \theta$. This transformation into the equation (\ref{b9}) leads to the following differential equation:
\begin{eqnarray}
    \psi''(\theta)+\Big(\frac{1}{\sin \theta\,\cos \theta}-\frac{2\,\sin \theta}{\cos \theta}\Big)\,\
    \psi' (\theta)+\Bigg[4\,\Lambda\,R^2-\frac{b^2\,\sin^2 \theta}{\cos^2 \theta}-\frac{j^2\,\cos^2 \theta}{\sin^2 \theta}\Bigg]\,\psi (\theta)=0\,.\label{b11}
\end{eqnarray}
By making two changes of variables, with the first one $x=\cos \theta$ and  the second one $\xi=1-x^2$ into the above Eq. (\ref{b11}) results
\begin{eqnarray}
    \xi\,(1-\xi)\,\psi''(\xi)+(1-2\,\xi)\,\psi'(\xi)+\Bigg[R^2\,\Lambda-\frac{b^2}{4}\,\frac{\xi}{1-\xi}-\frac{j^2}{4}\,\frac{(1-\xi)}{\xi} \Bigg]\,\psi(\xi)=0\,.\label{b12}
\end{eqnarray}

Let us require again that the solution must be finite at $\xi=0$ and $\xi=1$. We can write a possible solution to Eq. (\ref{b12}) as follows
\begin{eqnarray}
    \psi (\xi)=(1-\xi)^{\gamma}\,\xi^{\beta}\,F(\xi),\quad \beta=\frac{j}{2},\quad \gamma=\frac{b}{2}\,.\label{b13}
\end{eqnarray}
This new transformation leads to the following differential equation for the function 
\begin{eqnarray}
     \xi\,(1-\xi)\,F''(\xi)+\Bigg[1+2\,\beta-\Big(2+2\,\beta+2\,\gamma\Big)\,\xi\Bigg]\,F'(\xi)+\Big[R^2\,\Lambda-\gamma-\beta-2\,\gamma\,\beta\Big]\,F(\xi)=0.\label{b14}
\end{eqnarray}

We can identify the equations; the above equation coincides with the general form of the hypergeometric equation. Making an identification with a general hypergeometric function $F(\mathcal{A}, \mathcal{B}, \mathcal{C}, \xi)$ satisfying the equation \cite{MA}
\begin{equation}
    \xi\,(1-\xi)\,F''(\xi)+\Big[\mathcal{C}-(1+\mathcal{A}+\mathcal{B})\,\xi \Big]\,F'(\xi)-\mathcal{A}\,\mathcal{B}\,F(\xi)=0,\label{b15}
\end{equation}
where
\begin{eqnarray}
\mathcal{C}=1+2\,\beta,\quad\quad \mathcal{A}+\mathcal{B}=1+2\,\beta+2\,\gamma,\quad\quad \mathcal{A}\,\mathcal{B}=-R^2\,\Lambda+\beta+\gamma+2\,\gamma\,\beta.\label{b16}
\end{eqnarray}
These two parameter $(\mathcal{A}, \mathcal{B})$ are given by
\begin{eqnarray}
    (\mathcal{A}, \mathcal{B})=\frac{1}{2}\,\Big[1+2\,\beta+2\,\gamma \pm \sqrt{1+4\,\beta^2+4\,\gamma^2+4\,\Lambda\,R^2}\Big]\,.\label{b17} 
\end{eqnarray}

The solution $F(\xi)$ in Eq. (\ref{b15}) is a polynomial of degree $n$ obtained by the Frobenius method of power series expansion. Note that we require regularity of the wave function both at the origin and at infinity in the Eq. (\ref{b13}). The truncation of the power series expansion $F(\xi)$ is possible by imposing the following condition
\begin{equation}
    \mathcal{A}=-n\quad (n=0,1,2,3,...)\,.\label{b18}
\end{equation}
Simplification of the above relation results the following eigenvalue equation
\begin{equation}
    \Lambda\,R^2=(n^2+n+2\,\beta\,n+\beta)+(2\,n+1+2\,\beta)\,\gamma.\label{b19}
\end{equation}
Explicitly writing the above equation, we obtain
\begin{eqnarray}
&&\Rightarrow \frac{\mathcal{F}^2\,E^2-M^2}{\mathcal{G}^2}\,R^2-R^2\,k^2-\frac{2\,m\,\Omega\,R^2}{\alpha}\,\frac{\mathcal{F}}{\mathcal{G}}\,E-\frac{m^2}{2\,\alpha^2}=\Big(n^2+n+\frac{|m|}{\alpha}\,n+\frac{|m|}{2\,\alpha}\Big)\nonumber\\
&&+\frac{1}{2}\,\Big(2\,n+1+\frac{|m|}{\alpha}\Big)\,\Big|\frac{m}{\alpha}+4\,\Omega\,R^2\,E\,\frac{\mathcal{F}}{\mathcal{G}}\Big|.\label{b20}
\end{eqnarray}

The corresponding wave function is 
\begin{equation}
    \psi (\xi)=\mathcal{N}_{n, m}\,(1-\xi)^{\frac{1}{2}|\frac{m}{\alpha}+4\,\Omega\,R^2\,E\,\frac{\mathcal{F}}{\mathcal{G}}|}\,\xi^{\frac{|m|}{2\,\alpha}}\,F\Big(-n, n+1+\frac{|m|}{\alpha}+\Big|\frac{m}{\alpha}+4\,\Omega\,R^2\,E\,\frac{\mathcal{F}}{\mathcal{G}}\Big|, 1+\frac{|m|}{\alpha}, \xi\Big),\label{wave-function}
\end{equation}
where the constants $\mathcal{N}_{n, m}$ are obtained from the condition of the normalization of the wave function.

We consider two scenario of the above eigenvalue equation: The first one is that $\Big(\frac{m}{\alpha}+4\,\Omega\,R^2\,E\,\frac{\mathcal{F}}{\mathcal{G}}\Big)>0$. In that case, from Eq. (\ref{b20}) we obtain
\begin{eqnarray}
    &&\Rightarrow \frac{\mathcal{F}^2\,E^2-M^2}{\mathcal{G}^2}\,R^2-R^2\,k^2-\frac{2\,m\,\Omega\,R^2}{\alpha}\,\frac{\mathcal{F}}{\mathcal{G}}\,E-\frac{m^2}{2\,\alpha^2}=\Big(n^2+n+\frac{|m|}{\alpha}\,n+\frac{|m|}{2\,\alpha}\Big)\nonumber\\
    &&+\frac{1}{2}\,\Big(2\,n+1+\frac{|m|}{\alpha}\Big)\,\Big(\frac{|m|}{\alpha}+4\,\Omega\,R^2\,E\,\frac{\mathcal{F}}{\mathcal{G}}\Big).\label{b21}
\end{eqnarray}

The second one is $\Big(\frac{m}{\alpha}+4\,\Omega\,R^2\,E\,\frac{\mathcal{F}}{\mathcal{G}}\Big) <0$. In that case, from Eq. (\ref{b20}) we obtain
\begin{eqnarray}
    &&\Rightarrow \frac{\mathcal{F}^2\,E^2-M^2}{\mathcal{G}^2}\,R^2-R^2\,k^2-\frac{2\,m\,\Omega\,R^2}{\alpha}\,\frac{\mathcal{F}}{\mathcal{G}}\,E-\frac{m^2}{2\,\alpha^2}=\Big(n^2+n+\frac{|m|}{\alpha}\,n+\frac{|m|}{2\,\alpha}\Big)\nonumber\\
    &&-\frac{1}{2}\,\Big(2\,n+1+\frac{|m|}{\alpha}\Big)\,\Big(\frac{-|m|}{\alpha}+4\,\Omega\,R^2\,E\,\frac{\mathcal{F}}{\mathcal{G}}\Big).\label{b22}
\end{eqnarray}

Below, we choose two pair of rainbow functions and obtain analytical expressions of the energy eigenvalue of the scalar particles within this spherical symmetrical G\"{o}del-type metric with topological defects.

\subsection{Eigenvalue solutions with rainbow function $\mathcal{F}=1$ and $\mathcal{G}=1+\kappa\,\chi$}

To obtain bound-states eigenvalue solution of the scalar particles using the eigenvalue Eqs. (\ref{b21})--(\ref{b22}), we choose the following pair of rainbow function given by \cite{AFG}
\begin{equation}
    \mathcal{F}=1,\quad \mathcal{G}=1+\kappa\,\chi,\quad \chi=\frac{|E|}{E_p},\label{c1} 
\end{equation}
where $\kappa$ is the rainbow parameter. 

Thereby, substituting this pair of rainbow function into the eigenvalue equation (\ref{b21}), we obtain
\begin{equation}
    (E^2-M^2)\,R^2=\Delta^2\,\Big(1+\frac{\kappa}{E_p}\,|E|\Big)^2+E\,\Big(1+\frac{\kappa}{E_P}\,|E|\Big)\,\Theta,\label{c2}
\end{equation}
where
\begin{equation}
    \Delta^2=R^2\,k^2_{z}+\Big(n+\frac{|m|}{\alpha}\Big)\,\Big(n+\frac{|m|}{\alpha}+1\Big),\quad \Theta=2\,\Omega\,R^2\,\Big(2\,n+1+\frac{m}{\alpha}+\frac{|m|}{\alpha}\Big).\label{c3}
\end{equation}

For $|E|=E$, simplification of the above relation (\ref{c2}) gives us the energy eigenvalue expression given by
\begin{eqnarray}
    E^{+}_{n,m}=\frac{\frac{2\,\kappa\,\Delta^2}{E_p}+\Theta +\sqrt{4\,\Delta^2\,R^2+\Theta^2+4\,M^2\,R^2\,\Big(R^2-\frac{\Delta^2\,\kappa^2}{E^2_{p}}-\frac{\kappa\,\Theta}{E_p}\Big)}}{2\,\Big(R^2-\frac{\Delta^2\,\kappa^2}{E^2_{p}}-\frac{\kappa\,\Theta}{E_p}\Big)}.\label{c4}
\end{eqnarray}

We generate Figure 1 for the energy spectra $E^{+}_{n, m}$ of Eq. (\ref{c4}) with respect to $E_p/\kappa$ by choosing values of the angular speed $\Omega$, the cosmic string parameter $\alpha$, and the radial quantum number $n$. We see that the energy levels gradually decreases with increasing $E_p/\kappa$ and this trend shifted upward with increasing the values of $\Omega$ in fig. 1(a) and $n$ in fig 1(c) while shifted downward in fig. 1(b). 

In the limit $R \to \infty$, the space-time under consideration (\ref{b1}) becomes the Som-Raychaudhuri metric with topological defects. Therefore, the energy eigenvalue $E^{+}_{n, m}$ of scalar particles becomes 
\begin{eqnarray}
&&E^{+}_{n,m}=\frac{1}{{\Bigg[1-\frac{k^2_{z}\,\kappa^2}{E^2_{p}}-\frac{2\,\kappa\,\Big(2\,n+1+\frac{m}{\alpha}+\frac{|m|}{\alpha}\Big)\,\Omega}{E_p}\Bigg]}}\,\Bigg[\frac{\kappa\,k^2_{z}}{E_p}+\Big(2\,n+1+\frac{m}{\alpha}+\frac{|m|}{\alpha}\Big)\,\Omega\nonumber\\
&&+\sqrt{k^2_{z}+\Big(2\,n+1+\frac{m}{\alpha}+\frac{|m|}{\alpha}\Big)^2\,\Omega^2+M^2\,\Bigg\{1-\frac{k^2_{z}\,\kappa^2}{E^2_{p}}-\frac{2\,\kappa\,\Big(2\,n+1+\frac{m}{\alpha}+\frac{|m|}{\alpha}\Big)\,\Omega}{E_p}\Bigg\}}\Bigg].\label{c4c}
\end{eqnarray}
Note that in the limit $\kappa \to 0$, the energy eigenvalues Eq. (\ref{c4c}) reproduce the energy spectra $E^{+}$ of scalar particles obtained in the background of the Som–Raychaudhuri geometry with a cosmic string.

\begin{center}
\begin{figure}
\begin{centering}
\subfloat[$n=1,\alpha=0.5,m=1$]{\centering{}\includegraphics[scale=0.5]{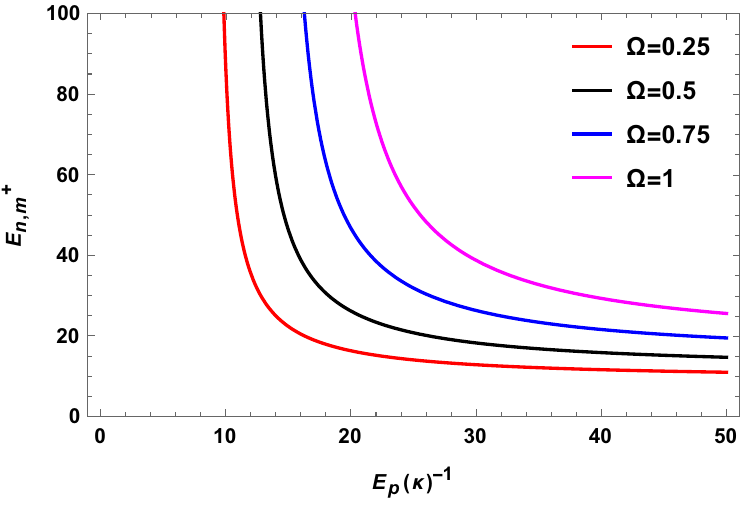}}\quad\quad
\subfloat[$n=1, \varOmega=0.5, m=1$]{\centering{}\includegraphics[scale=0.5]{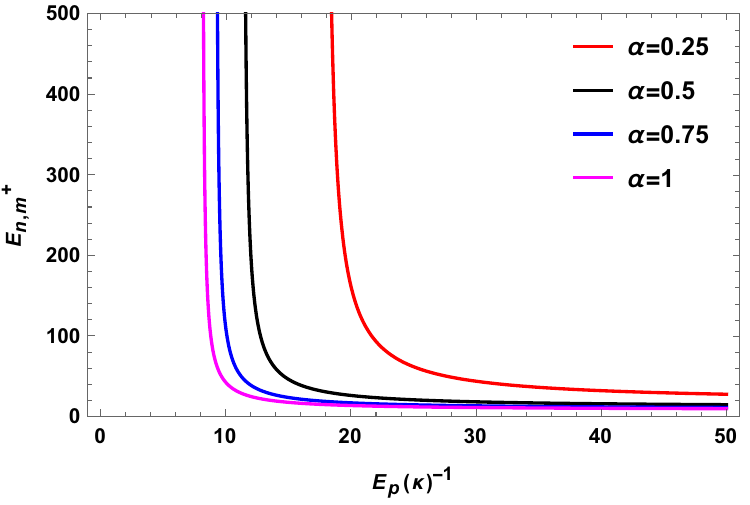}}
\par\end{centering}
\begin{centering}
\subfloat[$m=1,\varOmega=\alpha=0.5$]{\centering{}\includegraphics[scale=0.5]{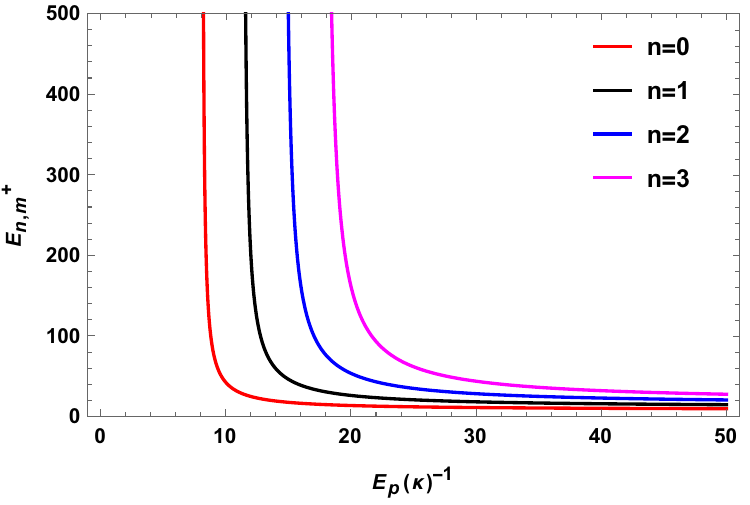}}
\par\end{centering}
\caption{Spectrum energy $E_{n,m}^{+}$ in relation (\ref{c4}), where $R=0.5, k_z=M=1$}
\end{figure}
\par\end{center}

For $|E|=-E$, simplification of the above relation (\ref{c2}) gives us the energy eigenvalue expression given by
\begin{eqnarray}
    E^{-}_{n,m}=\frac{-\frac{2\,\kappa\,\Delta^2}{E_p}+\Theta-\sqrt{4\,\Delta^2\,R^2+\Theta^2+4\,M^2\,R^2\,\Big(R^2-\frac{\Delta^2\,\kappa^2}{E^2_{p}}+\frac{\kappa\,\Theta}{E_p}\Big)}}{2\,\Big(R^2-\frac{\Delta^2\,\kappa^2}{E^2_{p}}+\frac{\kappa\,\Theta}{E_p}\Big)}.\label{c5}
\end{eqnarray}

We generate Figure 2 for the energy spectra $E^{-}_{n, m}$ of Eq. (\ref{c5}) with respect to $E_p/\kappa$ by choosing values of the angular speed $\Omega$, the cosmic string parameter $\alpha$, and the radial quantum number $n$ and shows their behaviour with increasing the values of these parameters.

In the limit $R \to \infty$, the energy eigenvalue $E^{-}_{n, m}$ reduces to
\begin{eqnarray}
&&E^{-}_{n,m}=\frac{1}{{\Bigg[1-\frac{k^2_{z}\,\kappa^2}{E^2_{p}}+\frac{2\,\kappa\,\Big(2\,n+1+\frac{m}{\alpha}+\frac{|m|}{\alpha}\Big)\,\Omega}{E_p}\Bigg]}}\,\Bigg[-\frac{\kappa\,k^2_{z}}{E_p}+\Big(2\,n+1+\frac{m}{\alpha}+\frac{|m|}{\alpha}\Big)\,\Omega\nonumber\\
&&-\sqrt{k^2_{z}+\Big(2\,n+1+\frac{m}{\alpha}+\frac{|m|}{\alpha}\Big)^2\,\Omega^2+M^2\,\Bigg\{1-\frac{k^2_{z}\,\kappa^2}{E^2_{p}}+\frac{2\,\kappa\,\Big(2\,n+1+\frac{m}{\alpha}+\frac{|m|}{\alpha}\Big)\,\Omega}{E_p}\Bigg\}}\Bigg].\label{c5c}
\end{eqnarray}
Note that in the limit $\kappa \to 0$, the energy eigenvalues Eq. (\ref{c5c}) reproduce the energy spectra $E^{-}$ of scalar particles obtained in the background of the Som–Raychaudhuri geometry with a cosmic string.

\begin{center}
\begin{figure}
\begin{centering}
\subfloat[$n=1,\alpha=0.5,m=1$]{\centering{}\includegraphics[scale=0.5]{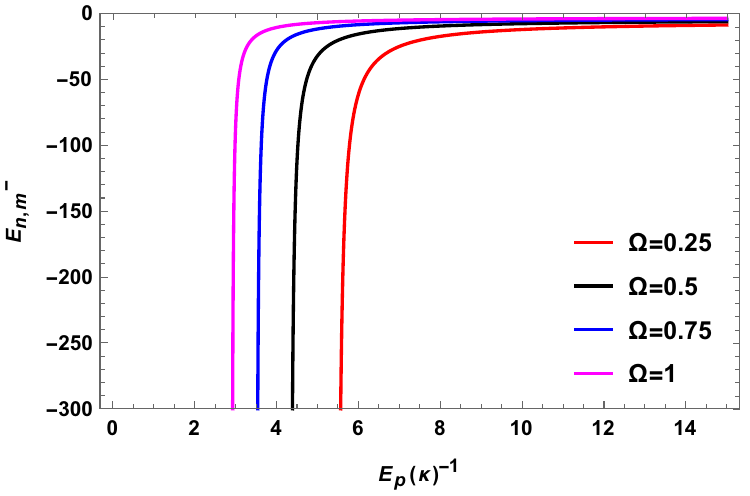}}\quad\quad
\subfloat[$n=1,\varOmega=0.5,m=1$]{\centering{}\includegraphics[scale=0.5]{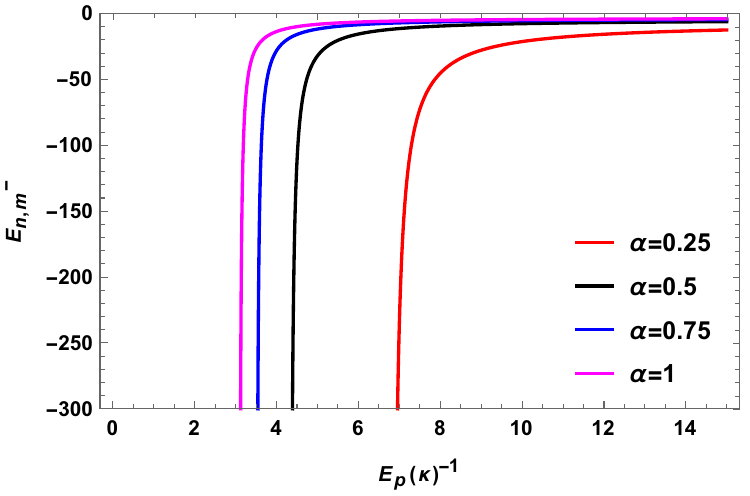}}
\par\end{centering}
\begin{centering}
\subfloat[$m=1,\varOmega=\alpha=0.5$]{\centering{}\includegraphics[scale=0.5]{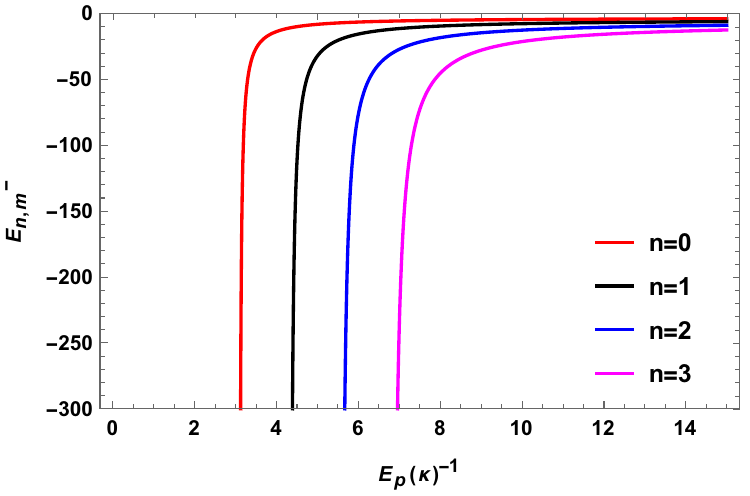}}
\par\end{centering}
\caption{Spectrum energy $E_{n,m}^{-}$ in relation (\ref{c5}), where $R=0.5, k_z=M=1$}
\end{figure}
\par\end{center}

Moreover, we discuss a special case corresponds to $\kappa \to 0$. In that limit, both the rainbow function (\ref{c1})becomes unity. Therefore, the quantum system under investigation is free from the effects of rainbow gravity's. Therefore, the energy eigenvalue expression Eqs. (\ref{c4}), (\ref{c5}) reduces to the following form
\begin{eqnarray}
    E^{\pm}_{n,m}=\frac{\Theta \pm \sqrt{4\,\Delta^2\,R^2+\Theta^2+4\,M^2\,R^4}}{2\,R^2}.\label{cc5}
\end{eqnarray}
Substituting $\Delta^2$ and $\Theta$ from Eq. (\ref{c3}) into the above equation (\ref{cc5}) results the final energy eigenvalue expression given by
\begin{eqnarray}
&&E^{\pm}_{n,m}=\Big(2\,n+1+\frac{m}{\alpha}+\frac{|m|}{\alpha}\Big)\,\Omega\nonumber\\
&&\pm\sqrt{\Big(2\,n+1+\frac{m}{\alpha}+\frac{|m|}{\alpha}\Big)^2\,\Omega^2+\frac{1}{R^2}\Big(n+\frac{|m|}{\alpha}\Big)\Big(n+\frac{|m|}{\alpha}+1\Big)+M^2+k^2_{z}}.\label{ccc5}
\end{eqnarray}

The above energy eigenvalue expression for scalar particles is similar to the outcome obtained in the study of spherical symmetrical G\"{o}del space-time with topological defects presented in Ref. \cite{MM29} (see Eq. (25) in Ref. \cite{MM29}). Therefore, a direct comparison between the expressions in Eqs. (\ref{c4}), (\ref{c5}) and the previous results (\ref{ccc5}), considering the presence of the rainbow parameter $\kappa$, reveals notable modifications in the energy levels of scalar particles. In Eqs. (\ref{c4}), (\ref{c5}), the influence of the rainbow parameter $\kappa$ is evident, signifying a departure from the earlier findings (\ref{ccc5}) established in the absence of rainbow gravity effects as reported in Ref. \cite{MM29} within the framework of spherical symmetrical G\"{o}del space-time with topological defects. This underscores the significance of the rainbow gravity contribution in shaping the energy spectrum of scalar particles, thereby introducing amendments to the previously known results in the context of curved space-time with topological defects.

Again, substituting this pair of rainbow function (\ref{c1}) into the eigenvalue equation (\ref{b22}), we obtain
\begin{equation}
    (E^2-M^2)\,R^2=\Delta^2\,\Big(1+\frac{\beta}{E_p}\,|E|\Big)^2+E\,\Big(1+\frac{\beta}{E_P}\,|E|\Big)\,\tilde{\Theta},\label{c6}
\end{equation}
where $\Delta^2$ is given in Eq. (\ref{c3}) earlier and 
\begin{equation}
    \tilde{\Theta}=2\,\Omega\,R^2\,\Big(2\,n+1+\frac{|m|}{\alpha}-\frac{m}{\alpha}\Big).\label{c7}
\end{equation}

For $|E|=E$, simplification of the above relation (\ref{c6}) gives us the energy eigenvalue expression given by
\begin{eqnarray}
  E^{+}_{n,m}=\frac{\frac{2\,\kappa\,\Delta^2}{E_p}+\tilde{\Theta}+\sqrt{4\,\Delta^2\,R^2+\tilde{\Theta}^2+4\,M^2\,R^2\,\Big(R^2-\frac{\Delta^2\,\kappa^2}{E^2_{p}}-\frac{\kappa\,\tilde{\Theta}}{E_p}\Big)}}{2\,\Big(R^2-\frac{\Delta^2\,\kappa^2}{E^2_{p}}-\frac{\kappa\,\tilde{\Theta}}{E_p}\Big)}.\label{c8}
\end{eqnarray}

We generate Figure 3 for the energy spectra $E^{+}_{n, m}$ of Eq. (\ref{c8}) with respect to $E_p/\kappa$ by choosing values of the angular speed $\Omega$, and the radial quantum number $n$ and shows their behaviour with increasing the values of these parameters.

\begin{center}
\begin{figure}
\begin{centering}
\subfloat[$n=1,\alpha=0.5,m=1$]{\centering{}\includegraphics[scale=0.5]{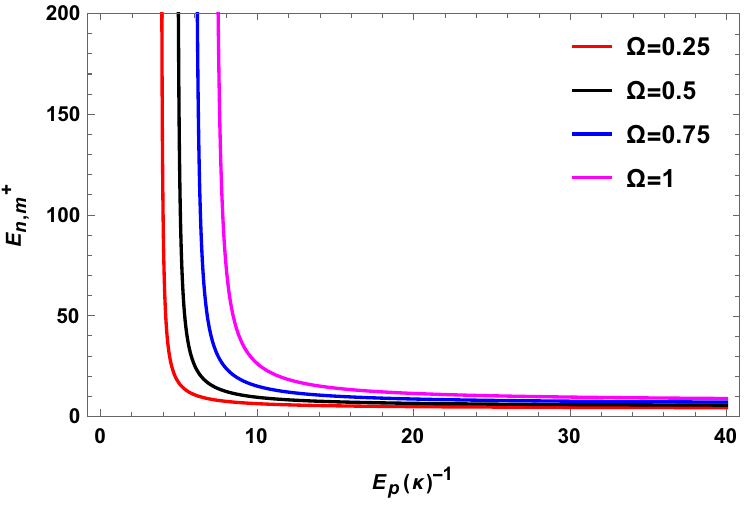}}\quad\quad
\subfloat[$\varOmega=\alpha=0.5,m=1$]{\centering{}\includegraphics[scale=0.5]{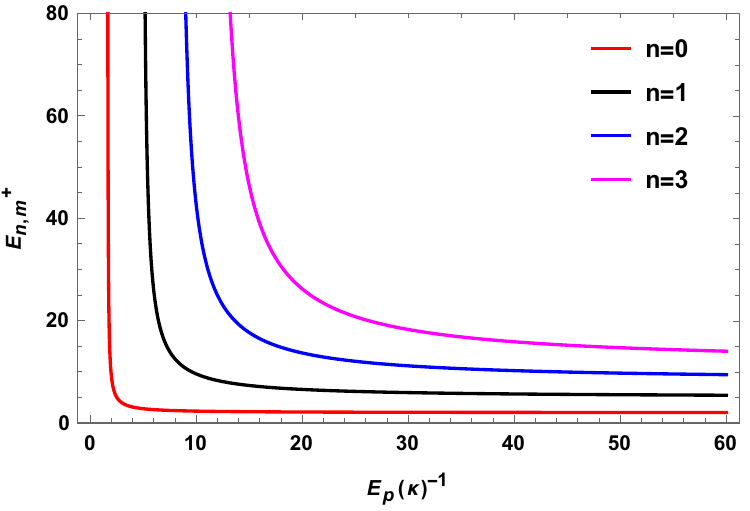}}
\par\end{centering}
\caption{Spectrum energy $E_{n,m}^{+}$ in relation (\ref{c8}), where $R=0.5, k_z=M=1$}
\end{figure}
\par\end{center}

For $|E|=-E$, simplification of the above relation (\ref{c6}) gives us the energy eigenvalue expression given by
\begin{eqnarray}
   E^{-}_{n,m}=\frac{-\frac{2\,\kappa\,\Delta^2}{E_p}+\tilde{\Theta}-\sqrt{4\,\Delta^2\,R^2+\tilde{\Theta}^2+4\,M^2\,R^2\,\Big(R^2-\frac{\Delta^2\,\kappa^2}{E^2_{p}}+\frac{\kappa\,\tilde{\Theta}}{E_p}\Big)}}{2\,\Big(R^2-\frac{\Delta^2\,\kappa^2}{E^2_{p}}+\frac{\kappa\,\tilde{\Theta}}{E_p}\Big)}.\label{c9}
\end{eqnarray}

We generate Figure 4 for the energy spectra $E^{-}_{n, m}$ of Eq. (\ref{c9}) with respect to $E_p/\kappa$ by choosing values of the angular speed $\Omega$, the cosmic string parameter $\alpha$, and the radial quantum number $n$ and shows their behaviour with increasing the values of these parameters.

\begin{center}
\begin{figure}
\begin{centering}
\subfloat[$n=1,\alpha=0.5,m=1$]{\centering{}\includegraphics[scale=0.5]{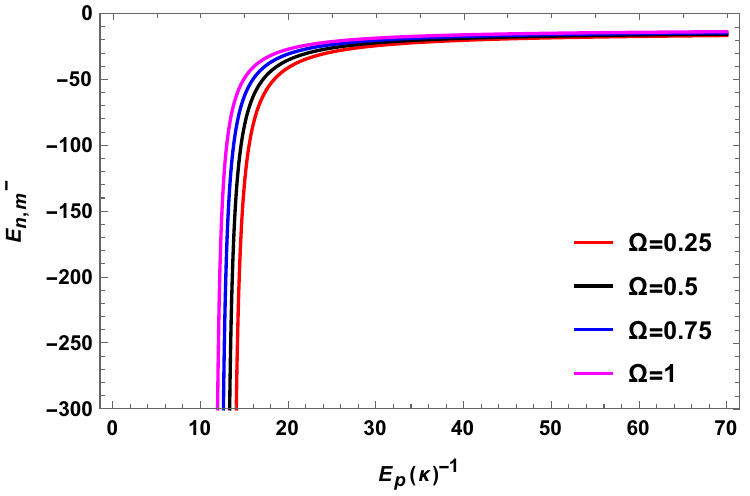}}\quad\quad
\subfloat[$\varOmega=\alpha=0.5,m=1$]{\centering{}\includegraphics[scale=0.5]{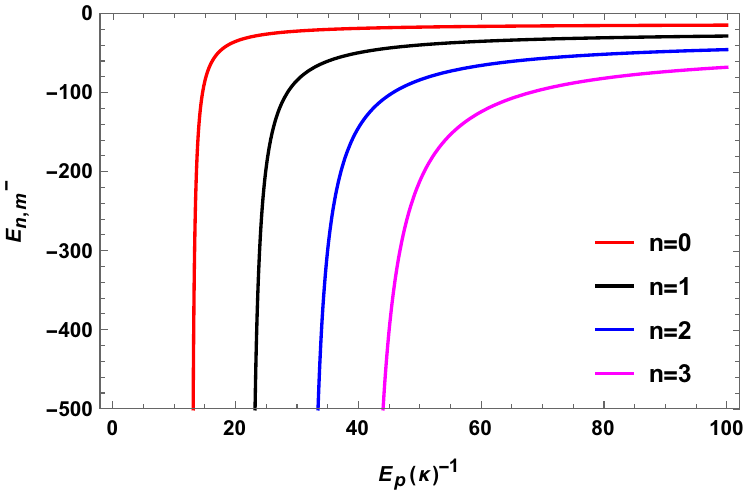}}
\par\end{centering}
\caption{Spectrum energy $E_{n,m}^{-}$ in relation (\ref{c9}), where $R=0.1, k_z=M=1$}
\end{figure}
\par\end{center}

In the limit $\kappa \to 0$, the chosen pair of rainbow function becomes unity. Therefore, the quantum system under investigation is now free from the influence of rainbow gravity's. Hence, using the expression Eqs. (\ref{c8})--(\ref{c9}), we obtain  
\begin{eqnarray}
  E^{\pm}_{n,m}=\frac{\tilde{\Theta}\pm\sqrt{4\,\Delta^2\,R^2+\tilde{\Theta}^2+4\,M^2\,R^4}}{2\,R^2}.\label{c10}
\end{eqnarray}
Substituting $\tilde{\Theta}$ and $\tilde{\Delta}^2$ from Eq. (\ref{c7}) into the above equation results
\begin{eqnarray}
E^{\pm}_{n,m}=\Big(2\,n+1+\frac{|m|}{\alpha}-\frac{m}{\alpha}\Big)\,\Omega\pm\sqrt{\Big(2\,n+1+\frac{|m|}{\alpha}-\frac{m}{\alpha}\Big)^2\Omega^2+\frac{1}{R^2}\,\Big(n+\frac{|m|}{\alpha}\Big)\,\Big(n+1+\frac{|m|}{\alpha}\Big)+M^2+k^2_{z}}.\label{c11}
\end{eqnarray}
Equation (\ref{c11}) is the relativistic energy eigenvalue of scalar particles in spherical symmetrical G\"{o}del-type metric with topological defect in the absence of rainbow gravity.

Moreover, in the limit $R \to \infty$, as stated earlier, the space-time under consideration becomes the Som-Raychaudhuri metric with topological defect. Therefore, the relativistic energy eigenvalue of scalar particle in the Som-Raychaudhuri space-time with topological defect in the presence of rainbow gravity's using the expression (\ref{c8}) after simplification becomes 
\begin{eqnarray}
&&E^{+}_{n,m}=\frac{1}{{\Bigg[1-\frac{k^2_{z}\,\kappa^2}{E^2_{p}}-\frac{2\,\kappa\,\Big(2\,n+1+\frac{|m|}{\alpha}-\frac{m}{\alpha}\Big)\,\Omega}{E_p}\Bigg]}}\,\Bigg[\frac{\kappa\,k^2_{z}}{E_p}+\Big(2\,n+1+\frac{|m|}{\alpha}-\frac{m}{\alpha}\Big)\,\Omega\nonumber\\
&&+\sqrt{k^2_{z}+\Big(2\,n+1+\frac{|m|}{\alpha}-\frac{m}{\alpha}\Big)^2\,\Omega^2+M^2\,\Bigg\{1-\frac{k^2_{z}\,\kappa^2}{E^2_{p}}-\frac{2\,\kappa\,\Big(2\,n+1+\frac{|m|}{\alpha}-\frac{m}{\alpha}\Big)\,\Omega}{E_p}\Bigg\}}\Bigg].\label{c12}
\end{eqnarray}
And that the energy eigenvalue $E^{-}_{n, m}$ using the expression (\ref{c9}) becomes
\begin{eqnarray}
&&E^{-}_{n,m}=\frac{1}{{\Bigg[1-\frac{k^2_{z}\,\kappa^2}{E^2_{p}}+\frac{2\,\kappa\,\Big(2\,n+1+\frac{|m|}{\alpha}-\frac{m}{\alpha}\Big)\,\Omega}{E_p}\Bigg]}}\,\Bigg[-\frac{\kappa\,k^2_{z}}{E_p}+\Big(2\,n+1+\frac{|m|}{\alpha}-\frac{m}{\alpha}\Big)\,\Omega\nonumber\\
&&+\sqrt{k^2_{z}+\Big(2\,n+1+\frac{|m|}{\alpha}-\frac{m}{\alpha}\Big)^2\,\Omega^2+M^2\,\Bigg\{1-\frac{k^2_{z}\,\kappa^2}{E^2_{p}}+\frac{2\,\kappa\,\Big(2\,n+1+\frac{|m|}{\alpha}-\frac{m}{\alpha}\Big)\,\Omega}{E_p}\Bigg\}}\Bigg].\label{c13}
\end{eqnarray}

\subsection{Eigenvalue solutions with rainbow function $\mathcal{F}=\frac{\Big(e^{\beta_0\,\chi}-1\Big)}{\beta_0\,\chi}$ and $\mathcal{G}=1$}


In this section, we choose the following pair of rainbow function given by \cite{GAC}
\begin{equation}
    \mathcal{F}=\frac{\Big(e^{\beta_0\,\chi}-1\Big)}{\beta_0\,\chi},\quad \mathcal{G}=1,\quad \chi=\frac{|E|}{E_p},\label{d1} 
\end{equation}
where $\beta_0$ is the rainbow parameter.

Substituting this pair of rainbow function into the eigenvalue equation (\ref{b21}), we obtain the following relation
\begin{equation}
    \frac{\Big(e^{\beta_0\,\chi}-1\Big)^2}{\beta^{2}_0\,\chi^2}\,E^2\,R^2-\frac{\Big(e^{\beta_0\,\chi}-1\Big)}{\beta_0\,\chi}\,E\,\Theta-(\Delta^2+M^2\,R^2)\,\Theta=0,\label{d2}
\end{equation}
where $\Theta$ and $\Delta^2$ are given in Eq. (\ref{c3}).

To obtain the energy levels, we first consider the case where $|E|=E$. Therefore, simplification of the above eigenvalue Eq. (\ref{d2}) under this case results the following expression of the relativistic energy eigenvalues given by
\begin{eqnarray}
&&E^{+}_{n, m}=\frac{1}{(\beta_0/E_p)}\,\ln\Bigg[1+\Bigg\{\Big(2\,n+1+\frac{m}{\alpha}+\frac{|m|}{\alpha}\Big)\,\Omega\nonumber\\
&&+\sqrt{\Big(2\,n+1+\frac{m}{\alpha}+\frac{|m|}{\alpha}\Big)^2\,\Omega^2+\frac{1}{R^2}\Big(n+\frac{|m|}{\alpha}\Big)\Big(n+\frac{|m|}{\alpha}+1\Big)+M^2+k^2_{z}}\Bigg\}\,(\beta_0/E_p)\Bigg].\label{d3}
\end{eqnarray}

We generate Figure 5 for the energy spectra $E^{-}_{n, m}$ of Eq. (\ref{d3}) with respect to $E_p/\beta_0$ by choosing values of the angular speed $\Omega$, the cosmic string parameter $\alpha$, and the quantum numbers $\{m, n\}$ and shows their behaviour with increasing the values of these parameters. We see that the energy levels almost linearly increases with increasing $E_p/\beta_0$ and this trend shifted upward with increasing values of the aforementioned parameters in Fig. 5 (a), (c)-(d) while shifted downward with increasing $\alpha$ in fig. 5(b).

\begin{center}
\begin{figure}
\begin{centering}
\subfloat[$n=1,\alpha=0.5,m=1$]{\centering{}\includegraphics[scale=0.5]{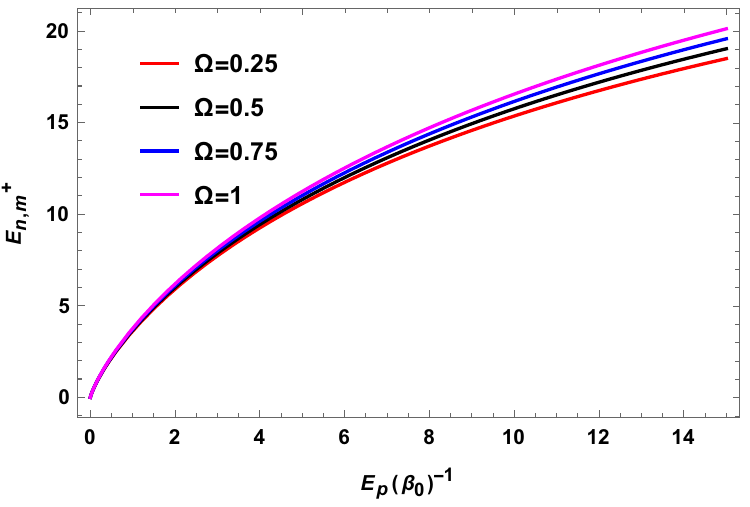}}\quad\quad
\subfloat[$\varOmega=0.5,n=m=1$]{\centering{}\includegraphics[scale=0.5]{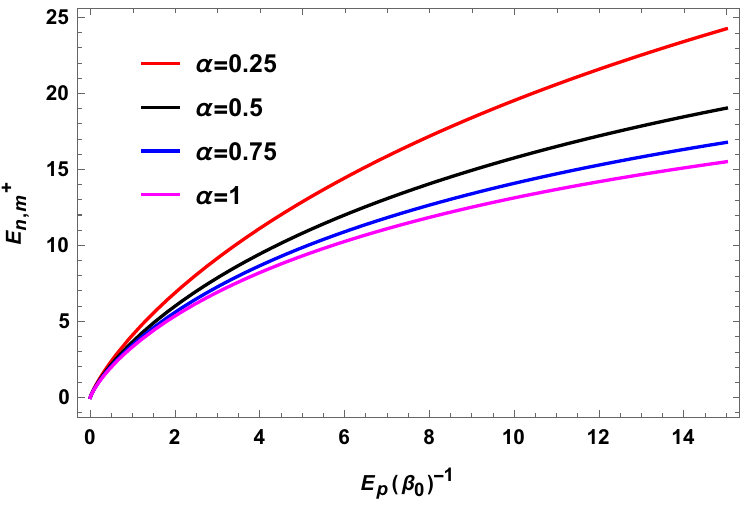}}
\par\end{centering}
\begin{centering}
\subfloat[$n=1,\alpha=\Omega=0.5$]{\centering{}\includegraphics[scale=0.5]{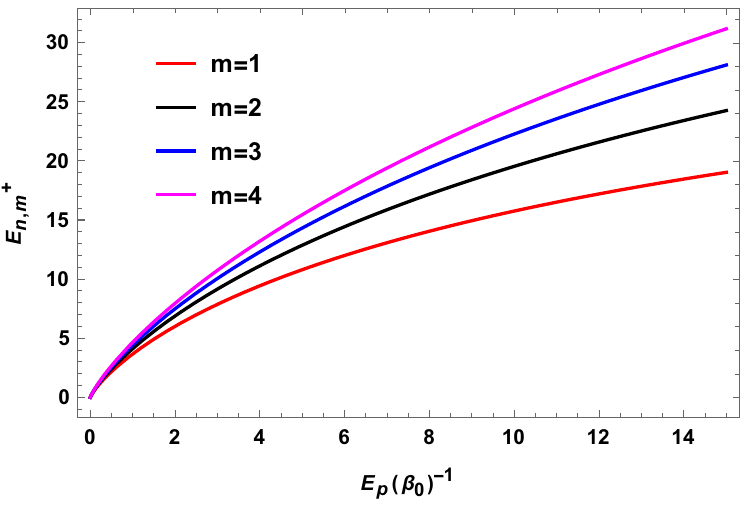}}\quad\quad
\subfloat[$\varOmega=\alpha=0.5,m=1$]{\centering{}\includegraphics[scale=0.5]{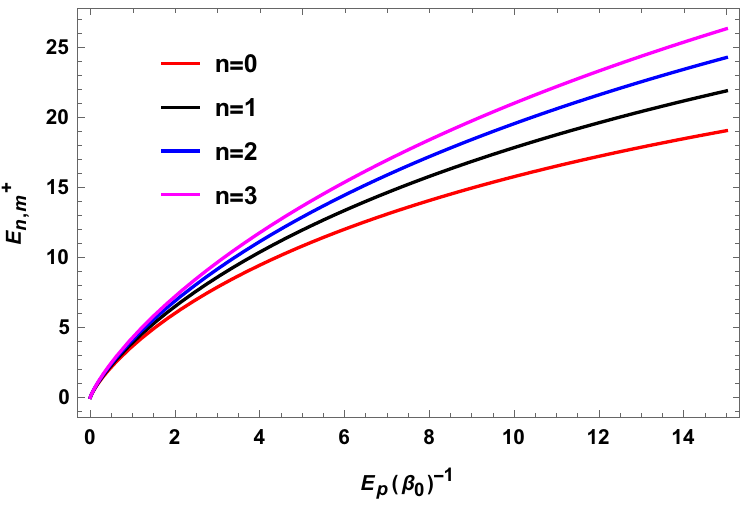}}
\par\end{centering}
\caption{Spectrum energy $E_{n,m}^{+}$ in relation (\ref{d3}), where $R=0.1, k_z=M=1$}
\end{figure}
\par\end{center}

Similarly, for $|E|=-E$, we obtain the following expression of the energy eigenvalues given by
\begin{eqnarray}
&&E^{-}_{n, m}=-\frac{1}{(\beta_0/E_p)}\,\ln\Bigg[1+\Bigg\{-\Big(2\,n+1+\frac{m}{\alpha}+\frac{|m|}{\alpha}\Big)\,\Omega \nonumber\\
&&+\sqrt{\Big(2\,n+1+\frac{m}{\alpha}+\frac{|m|}{\alpha}\Big)^2\,\Omega^2+\frac{1}{R^2}\Big(n+\frac{|m|}{\alpha}\Big)\Big(n+\frac{|m|}{\alpha}+1\Big)+M^2+k^2_{z}}\Bigg\}\,(\beta_0/E_p)\Bigg].\label{d4}
\end{eqnarray}

Equation (\ref{d3})--(\ref{d4}) is the relativistic energy eigenvalues $E^{\pm}$ of scalar particles in the spherical symmetrical G\"{o}del-type metric with topological defect in the presence of rainbow gravity chosen by the pair of function (\ref{d1}).

We generate Figure 6 for the energy spectra $E^{-}_{n, m}$ of Eq. (\ref{d4}) with respect to $E_p/\beta_0$ by choosing values of the angular speed $\Omega$, the cosmic string parameter $\alpha$, and the quantum numbers $\{n, m\}$ and shows their behaviour with increasing the values of these parameters.

\begin{center}
\begin{figure}
\begin{centering}
\subfloat[$n=1,\alpha=0.5,m=1$]{\centering{}\includegraphics[scale=0.5]{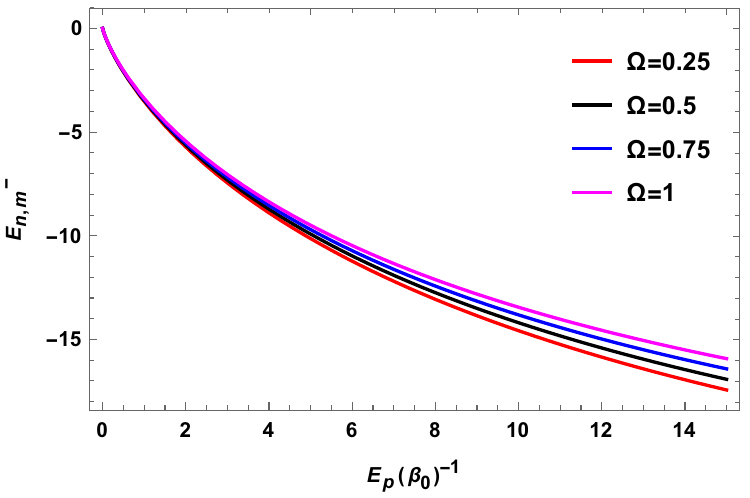}}\quad\quad
\subfloat[$\varOmega=0.5,n=m=1$]{\centering{}\includegraphics[scale=0.5]{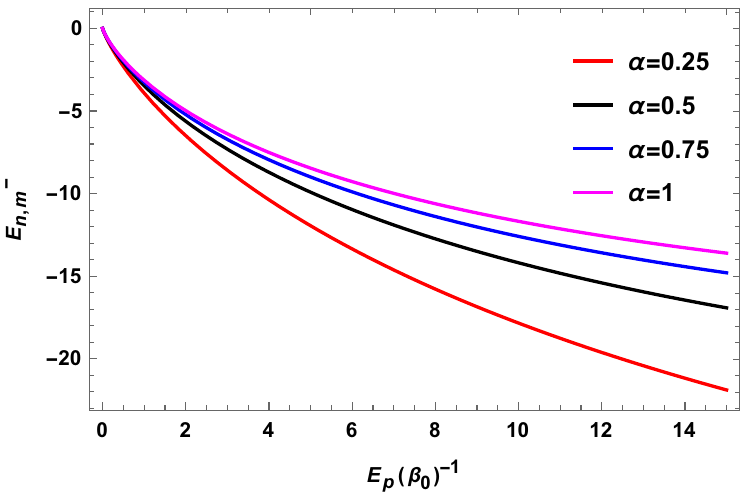}}
\par\end{centering}
\begin{centering}
\subfloat[$n=1,\alpha=\Omega=0.5$]{\centering{}\includegraphics[scale=0.5]{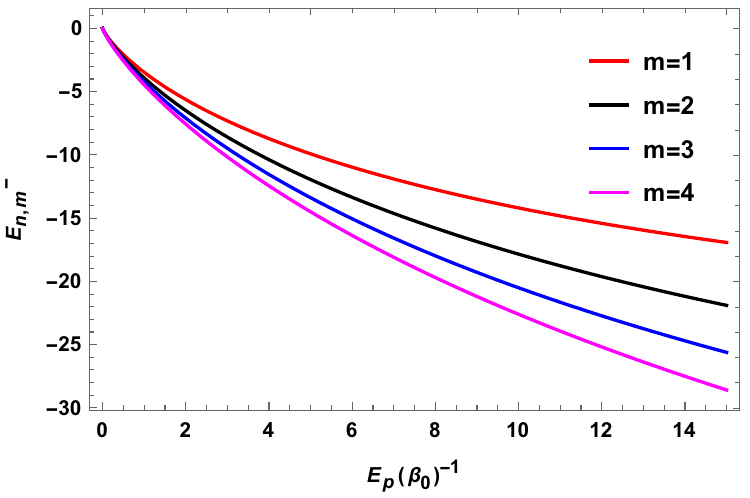}}\quad\quad
\subfloat[$\varOmega=\alpha=0.5,m=1$]{\centering{}\includegraphics[scale=0.5]{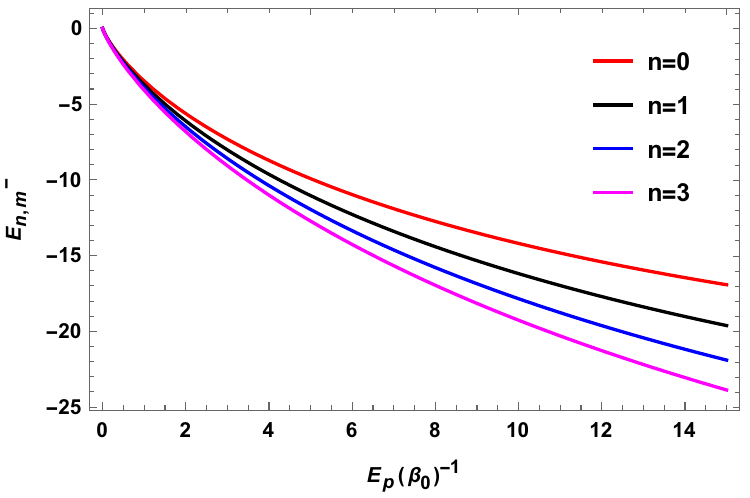}}
\par\end{centering}
\caption{Spectrum energy $E_{n,m}^{-}$ in relation (\ref{d4}), where $R=0.1, k_z=M=1$}
\end{figure}
\par\end{center}

In the limit $R \to \infty$, the space-time under consideration becomes the Som-Raychaudhuri metric with topological defect. Therefore, the relativistic energy eigenvalue of scalar particle in the Som-Raychaudhuri space-time with topological defect in the presence of rainbow gravity's using the expression (\ref{d3})--(\ref{d4}) becomes 
\begin{eqnarray}
E^{\pm}_{n, m}=\pm\frac{1}{(\beta_0/E_p)}\ln\Bigg[1+\Bigg\{\pm\,\Big(2n+1+\frac{m}{\alpha}+\frac{|m|}{\alpha}\Big)\Omega+\sqrt{\Big(2n+1+\frac{m}{\alpha}+\frac{|m|}{\alpha}\Big)^2\Omega^2+M^2+k^2_{z}}\Bigg\}(\beta_0/E_p)\Bigg].\label{d3d}
\end{eqnarray}

Again, substituting this pair of rainbow function (\ref{d1}) into the eigenvalue equation (\ref{b22}), we obtain
\begin{equation}
    \frac{\Big(e^{\beta_0\,\chi}-1\Big)^2}{\beta^{2}_0\,\chi^2}\,E^2\,R^2+\frac{\Big(e^{\beta_0\,\chi}-1\Big)}{\beta_0\,\chi}\,E\,\tilde{\Theta}-(\Delta^2+M^2\,R^2)=0,\label{d5}
\end{equation}
where $\tilde{\Theta}$ and $\Delta^2$ are given earlier. 

For $|E|=E$, after simplification of the above Eq. (\ref{d5}) gives the following expression:
\begin{eqnarray}
&&E^{+}_{n, m}=\frac{1}{(\beta_0/E_p)}\,\ln\Bigg[1+\Bigg\{-\Big(2\,n+1+\frac{|m|}{\alpha}-\frac{m}{\alpha}\Big)\,\Omega\nonumber\\
&&+\sqrt{\Big(2\,n+1+\frac{|m|}{\alpha}-\frac{m}{\alpha}\Big)^2\,\Omega^2+\frac{1}{R^2}\,\Big(n+\frac{|m|}{\alpha}\Big)\Big(n+\frac{|m|}{\alpha}+1\Big)+M^2+k^2_{z}}\Bigg\}\,(\beta_0/E_p)\Bigg].\label{d6}
\end{eqnarray}
Similarly, for $|E|=-E$, we obtain the following energy eigenvalue expression given by
\begin{eqnarray}
&&E^{-}_{n, m}=-\frac{1}{(\beta_0/E_p)}\,\ln\Bigg[1+\Bigg\{\Big(2\,n+1+\frac{|m|}{\alpha}-\frac{m}{\alpha}\Big)\,\Omega\nonumber\\
&&-\sqrt{\Big(2\,n+1+\frac{|m|}{\alpha}-\frac{m}{\alpha}\Big)^2\,\Omega^2+\frac{1}{R^2}\,\Big(n+\frac{|m|}{\alpha}\Big)\Big(n+\frac{|m|}{\alpha}+1\Big)+M^2+k^2_{z}}\Bigg\}\,(\beta_0/E_p)\Bigg].\label{d7}
\end{eqnarray}
Equation (\ref{d6})--(\ref{d7}) is the relativistic energy eigenvalues $E^{\pm}$ of scalar particles in the spherical symmetrical G\"{o}del-type metric with topological defect in the presence of rainbow gravity chosen by the pair of function (\ref{d1}).

In the limit $R \to \infty$, the energy eigenvalues expression (\ref{d3})--(\ref{d4}) reduces to the following form
\begin{eqnarray}
E^{+}_{n, m}=\frac{E_p}{\beta_0}\ln\Bigg[1+\Bigg\{-\Big(2n+1+\frac{|m|}{\alpha}-\frac{m}{\alpha}\Big)\Omega
+\sqrt{\Big(2n+1+\frac{|m|}{\alpha}-\frac{m}{\alpha}\Big)^2\Omega^2+M^2+k^2_{z}}\Bigg\}\frac{\beta_0}{E_p}\Bigg].\label{d8}
\end{eqnarray}
\begin{eqnarray}
E^{-}_{n, m}=-\frac{E_p}{\beta_0}\ln\Bigg[1+\Bigg\{\Big(2n+1+\frac{|m|}{\alpha}-\frac{m}{\alpha}\Big)\Omega
-\sqrt{\Big(2n+1+\frac{|m|}{\alpha}-\frac{m}{\alpha}\Big)^2\Omega^2+M^2+k^2_{z}}\Bigg\}\frac{\beta_0}{E_p}\Bigg].\label{d9}
\end{eqnarray}
Equation (\ref{d8})--(\ref{d9}) is the relativistic energy eigenvalues $E^{\pm}_{n, m}$ of scalar particles in the Som-Raychaudhuri metric with topological defect in the presence of rainbow gravity chosen by the pair of function (\ref{d1}).

\section{Conclusions}

In this contribution, we conducted an examination of the relativistic quantum dynamics governing spin-0 scalar particles, as described by the wave equation of the Klein-Gordon equation, within the backdrop of a curved space-time. Specifically, we explored a spherical symmetrical G\"{o}del-type metric in the presence of a topological defect generated by a cosmic string. To further extend our analysis, we introduced the influence of rainbow gravity's function into this curved space-time, thereby deforming the space-time geometry and altering the curvature properties due to energy-dependent effects. In our investigation, we considered a few noteworthy pair of rainbow function, the first pair is given by $\mathcal{F}=1$ and $\mathcal{G}=(1+\kappa\,\chi)$. This pair of function has been previously employed in studies related to the time delay of light signals \cite{AFG}. The second pair of rainbow function is $\mathcal{F}=\frac{(e^{\beta_0\,\chi}-1)}{\beta_0\,\chi}$ and $\mathcal{G}=1$, which has been utilized in testing of quantum gravity through observations of $\gamma$-ray bursts \cite{GAC}. Our results showed that the energy eigenvalues and the wave functions of scalar particles exhibit a dependence on the presence of the topological defect, consequently modifying the outcomes. Notably, the introduction of the topological defect leads to a breaking of the degeneracy observed in the energy spectra. Furthermore, we observed that the eigenvalue solutions of scalar particles are notably influenced by the rainbow parameter, specifically for the chosen pairs of rainbow function. This highlights the intricate interplay between the rainbow gravity effects and the underlying space-time geometry, showcasing the impact on the quantum properties of scalar particles. Our comprehensive analysis provides valuable insights into the intricate dynamics of relativistic quantum systems in the presence of both topological defects and rainbow gravity effects. To illustrated these energy spectra, we have generated Figures 1--6 by increasing values of the parameters, such as the constant angular speed $(\Omega)$, the cosmic string parameter $(\alpha)$, and the quantum numbers $\{n, m\}$ and showed their behaviors with increasing values of these parameters.

The introduction of a topological defect induces modification to the curvature properties of curved spaces. This alteration significantly also influences the behavior of bosonic and fermionic particles, imparting changes to their energy levels and wave functions when solved relativistic and non-relativistic wave equations. Furthermore, the integration of rainbow gravity into the geometry further enriches our comprehension by accounting for energy-dependent effects near the Planck's scale which modified or deformed the space-time under consideration. This methodology not only propels advancements in theoretical physics but also have potential applications in the context of cosmology and astrophysics. By bridging the realms of quantum mechanics and gravitational theory, our study offers a valuable framework for exploring the intricacies of cosmic phenomena and deepening our understanding of connection between these theories. 

Looking ahead, future endeavors in this field aim to tackle additional challenges and broaden the scope of study to encompass other space-time geometries, thus deepening our comprehension of curved space-time and topological defects in the presence of rainbow gravity's. By extending our analysis to diverse space-time backgrounds, we can valuable insights into the behavior of quantum particles in varied gravitational environments. Furthermore, exploring the implications of rainbow gravity in different space-time configurations holds promise for refining our understanding of quantum gravity effects and their interplay with topology.







\end{document}